\documentclass{emulateapj}

\usepackage{natbib}
\usepackage{hyperref}

\shorttitle{Fading hard X-ray emission from Sgr B2}
\shortauthors{Terrier et al.}

\begin{document}

\title{Fading hard X-ray emission from the Galactic Centre molecular cloud Sgr B2}

\author{R. Terrier$^1$, G. Ponti$^{1,2}$, G. B\'elanger$^3$, A. Decourchelle$^{4,5}$, V. Tatischeff$^6$, A. Goldwurm$^{1,4}$ , G. Trap$^{1,4}$, M. R. Morris$^{7}$ and R. Warwick$^{8}$ }

\affil{$^1$Astroparticule et Cosmologie, Universit\'e Paris7/CNRS/CEA, Batiment Condorcet, 75013 Paris, France}
\affil{$^2$School of Physics and Astronomy, University of Southampton, Highfield, Southampton, SO17 1BJ, UK}
\affil{$^3$European Space Agency, ESAC, P.O. Box 78, 28691,  Villanueva de la Canada (Madrid), Spain}
\affil{$^4$Service d'Astrophysique (SAp)/IRFU/DSM/CEA Saclay, Bt. 709, 91191 Gif-sur-Yvette Cedex, France}
\affil{$^5$Laboratoire AIM, CEA-IRFU/CNRS/Universit\'e Paris Diderot, CEA Saclay, 91191 Gif sur Yvette, France}
\affil{$^6$Centre de Spectrom\'etrie Nucl\'eaire et de Spectrom\'etrie de Masse, CNRS/IN2P3 and Univ Paris-Sud, 91405 Orsay, France}
\affil{$^7$Department of Physics \& Astronomy, University of California, Los Angeles, CA 90095-1547, USA}
\affil{$^8$Department of Physics \& Astronomy, University of Leicester, University Road, Leicester LE1 7RH, UK}

\email{rterrier@in2p3.fr}

\begin{abstract}
The centre of our Galaxy harbours a 4 million solar mass black hole that is unusually quiet: its present X-ray luminosity  is more than 10 orders of magnitude less than its Eddington luminosity. The observation of iron fluorescence and hard X-ray emission from some of the massive molecular clouds surrounding the Galactic Centre has been interpreted as an echo of a past $\mathrm{10^{39}\ erg\ s^{-1}}$ flare. Alternatively, low-energy cosmic rays propagating inside the clouds might account for the observed emission, through inverse bremsstrahlung of low energy ions or bremsstrahlung emission of low energy electrons. Here we report the observation of a clear decay of the hard X-ray emission from the molecular cloud Sgr B2 during the past 7 years thanks to more than 20 Ms of INTEGRAL exposure.  This confirms the decay previously observed comparing the 6.4 keV line fluxes measured by various X-ray instruments, but without intercalibration effects. The measured decay time is 8.2 $\pm$ 1.7 years, compatible with the light crossing time of the molecular cloud core . Such a short timescale rules out inverse bremsstrahlung by cosmic-ray ions as the origin of the X ray emission.
We also obtained 2-100 keV broadband X-ray spectra by combining INTEGRAL and  XMM-Newton data and compared them with detailed models of X-ray emission due  to irradiation of molecular gas by (i) low-energy cosmic-ray electrons and (ii) hard X-rays. Both models can reproduce the data equally well, but the time variability constraints and the huge cosmic ray electron luminosity required to explain the observed hard X-ray emission strongly favor the scenario in which the diffuse emission of Sgr B2 is scattered and reprocessed radiation emitted in the past by Sgr A*. The spectral index of the illuminating power-law source is found to be $\Gamma \sim 2$ and its luminosity $\mathrm{1.5-5 \times 10^{39}\ erg\ s^{-1}}$, depending on the relative positions of Sgr B2 and Sgr A$^*$.  Using recent parallax measurements that place Sgr B2 in front of Sgr A$^*$, we find that the period of intense activity of Sgr A$^*$ ended between 75 and 155 years ago.

\end{abstract}


\section{Introduction}

The supermassive black hole of $\mathrm{4.3 \times 10^6\ M_\odot}$ \citep{Ghez08,Gillessen09} that sits at the centre of the Galaxy is remarkably quiescent nowadays with an X-ray luminosity of about $\mathrm{10^{33-34}\ erg s^{-1}}$, more than 10 orders of magnitude less than its Eddington luminosity. The brightest X-ray flare recorded so far reached a level of a few  $\mathrm{10^{35}\ erg s^{-1}}$ \citep{Porquet08}. 
Active Galactic Nuclei have a relatively low duty cycle ($\mathrm{\sim 10^{-2}}$, \citealt{Ho08}), so that they spend most of the time  in a low luminosity state. It is therefore not excluded that Sgr A$^*$ has experienced periods of increased activity in the past. 
Looking for evidence of such a past activity of Sgr A$^*$, we turn to its surroundings that would probably carry the scars of such an event.

The central molecular zone or CMZ \citep{Morris96}, although relatively small in size with a radius of about 200 pc, contains about 10\% of the Galaxy's molecular mass. It is  populated by some of the most massive ($\mathrm{10^4-10^6\ M_{\odot}}$) and dense ($\mathrm{>\ 10^5\  cm^{-3}}$) giant molecular complexes in the Galaxy. 
Sgr B2 is the densest and most massive among them with a mass of a few $\mathrm{10^6 M_\odot}$, and density in its 5 pc core reaching $\mathrm{10^6\ cm^{-3}}$. It has an envelope that extends to $\sim$20 pc with an average density of $\sim\mathrm{10^{3}\ cm^{-3}}$  \citep{Goldsmith90}.  

The CMZ is also a source of non-thermal radiation. ART-P on board GRANAT was the first to detect hard X-ray emission from the region \citep{Sunyaev93}. The ASCA satellite discovered a strong iron K$\alpha$ fluorescence line at 6.4 keV with a large equivalent width of 1-2 keV from Sgr B2 \citep{Koyama96}. Chandra has surveyed the central region of our Galaxy and mapped the distribution of Fe line emission from the molecular clouds \citep{Park04}.
Today, thanks to the improved sensitivity of INTEGRAL/IBIS above 20 keV, the emission of Sgr B2 was resolved from the rest of the Galactic Centre (GC) sources \citep{Revnivtsev04}.

The features seen in Sgr B2 are not unique among clouds in the CMZ. G0.13--0.13 (also konwn as G0.11-0.11) displays strong 6.4 keV line emission \citep{Yusef02}, as well as a hard X-ray continuum detected at energies above 50 keV \citep{Belanger06}. Sgr C shows Fe K$\alpha$ line emission \citep{Nakajima09}, however its associated hard X-ray continuum cannot be resolved by current instruments because it lies close to a bright X-ray binary, KS 1741-293 \citep{deCesare07}.

The origin of this non-thermal emission is controversial.  A likely explanation for the continuum and the strong K$\alpha$ line, is  Compton scattering and K-shell photo-ionization of neutral or low-ionized iron atoms by an intense X-ray radiation field generated in the GC \citep{Sunyaev93, Koyama96, Sunyaev98}. Here, the clouds would act as X-Ray Reflection Nebulae (XRN).  Since there has been no report of a sufficiently bright X-ray transient (L $\ge 10^{37}$ erg/s) close to Sgr B2 in the last 20 years, it has been argued that the origin of the emission should lie in the most central regions and have a luminosity of at least $ 10^{39}$ erg/s for several years. The most probable candidate  is Sgr A$^*$. This, however, requires an increase of more than 6 orders of magnitude compared to its current luminosity\citep{Koyama96}. It has also been suggested that the Sgr A East supernova remnant interacting with the  GC massive molecular clouds could provide the required luminosity\citep{Fryer06}.

The propagation of energetic charged particles inside the molecular clouds has also been invoked as an explanation of the observed X-rays from the region. For instance, \citet{Yusef07} suggested that the density of cosmic rays (CR) is larger in the CMZ than in the rest of the Galaxy. Their hypothesis is based on several results: the observation of diffuse low-frequency radio emission in the central regions \citep{LaRosa05},  the high estimates of the ionization rate compared to the values obtained in the Galactic disk \citep{Oka05}, and the detection of hard interstellar emission from the molecular clouds at TeV energies \citep{Aharonian06}.

Two distinct processes can produce hard X-ray continuum  and neutral iron line emissions.
First, low-energy cosmic-ray electrons (LECRE), with energies up to a few hundred keV, diffusing into dense neutral matter will produce nonthermal bremsstrahlung and collisional K-shell ionization \citep{Valinia00}. The presence of strong non-thermal radio filaments in the vicinity of several molecular clouds makes it reasonable to consider the presence of a large flux of cosmic ray electrons in these regions, see e.g. G0.13-0.13 \citep{Yusef02} or Sgr C \citep{Yusef07}.  
Second, subrelativistic ions propagating in the molecular cloud can radiate through inverse bremsstrahlung in the hard X-ray domain and create iron K$\alpha$ vacancies. Knock-on electrons generated by collisions of primary ions with the gas also contribute significantly to the total emission \citep{Dogiel09}.  These subrelativistic ions can also be invoked to explain the putative hot plasma pervading the GC. 

 \citet{Revnivtsev04} have argued that the X-ray emission from Sgr B2 is difficult to explain by LECREs, because the large power that 
cosmic-ray particles would need to deposit in the molecular gas to account for the X-ray flux would be comparable to the bolometric (mostly 
infrared) luminosity of the complex. They also put into question the large metallicities implied by the large equivalent width of the line produced by collisional K-shell interaction.

To help resolve this ambiguity, Sgr B2 and the other clouds of the CMZ have been the subject of numerous observation campaigns.
In particular, a deep Chandra observation revealed that the 6.4 keV line emission is oriented towards the GC \citep{Murakami01}. It also allowed the contribution from individual sources in the GMC associated with dust condensations (e.g. Sgr B2 North (N) Middle (M) and South(S) to be distinguished \citealp{Takagi02}).

Recently, thanks to the monitoring of the region by many instruments, time dependent effects have been observed. First, \citet{Muno07}
 have detected time varying diffuse emission in the 4-8 keV range a few arcmin away from  Sgr A$^{\star}$  comparing Chandra images taken before and after 2004. Since this region is emitting the 6.4 keV line, it is likely that the varying diffuse emission can be linked with iron fluorescence, making this cloud  a likely  X-ray reflection nebula.  More recently, comparing recent observations with various X-ray telescopes (namely Suzaku, XMM-Newton, Chandra), \citet{Inui09} were able to show a temporal decrease in the K$\alpha$ line emission from Sgr B2, strengthening the reflection interpretation for this cloud as well. 

The hard X-ray continuum carries many clues as regards the origin of the emission: its spectral shape and its temporal behaviour in particular. 
With its sensitivity and with its very long exposures of the Galactic central region, INTEGRAL can therefore provide invaluable insight into the nature of the hard X-ray emission of the CMZ. We present here the lightcurve and spectrum of the Sgr B2 emission taken over 7 years of observations and show that the Sgr B2 flux has decreased by more than 40\% since 2003. We then study the combined XMM-Newton/INTEGRAL spectrum obtained from observations prior to 2004 which we use to test LECRE and XRN models. Finally, we discuss the implications of our results in relation to the likely nature of the non thermal emission from the GC molecular clouds.

\section{Observations and data analysis}

\begin{figure*}
\epsscale{1}
\plotone{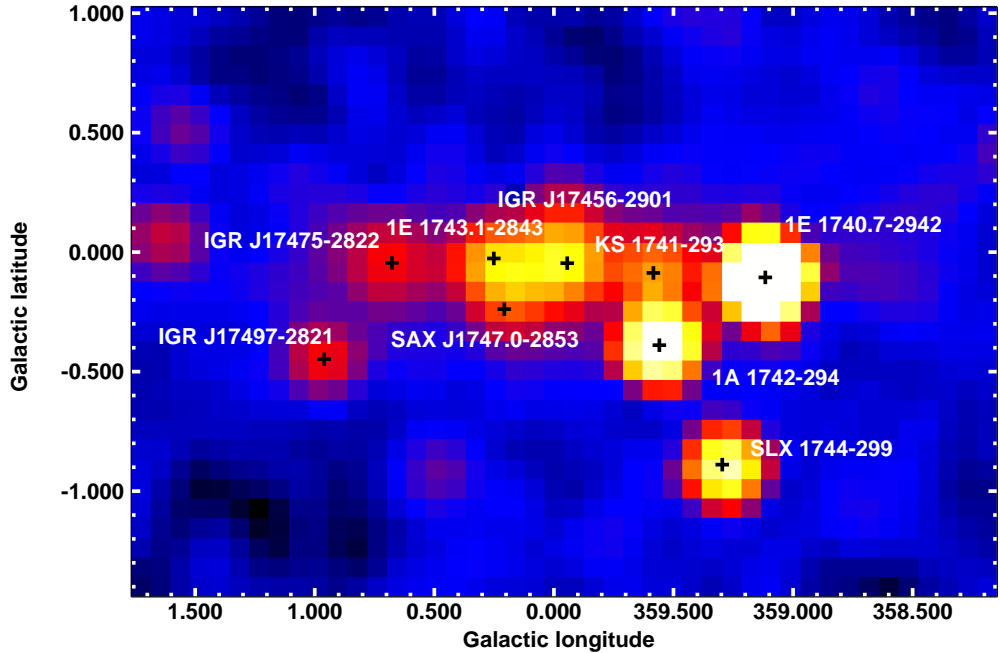}
\caption{IBIS/ISGRI significance map of the Galactic central regions in the 20-40 keV energy range. The image was produced combining  all the GC observations. The main sources are indicated by the black crosses. The position of IGR J17475-2822 is compatible with that of the Sgr B2 molecular cloud.\label{fig:IsgriImage}}
\end{figure*}

\begin{figure}
\epsscale{1.2}
\plotone{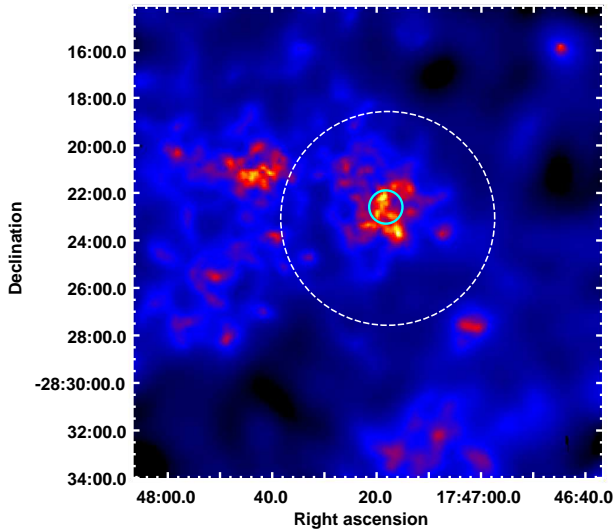}
\caption{XMM-Newton/MOS 6.4 keV line flux image of the Sgr B2 field. The cyan circle gives the best fit position of  IGR J17475-2822 with the associated 90\% confidence error on the position. The white dashed circle shows the 4.5' wide extraction region used for the combined spectral analysis.\label{fig:IsgriPos}}
\end{figure}

\begin{figure*}
\epsscale{0.55}
\plotone{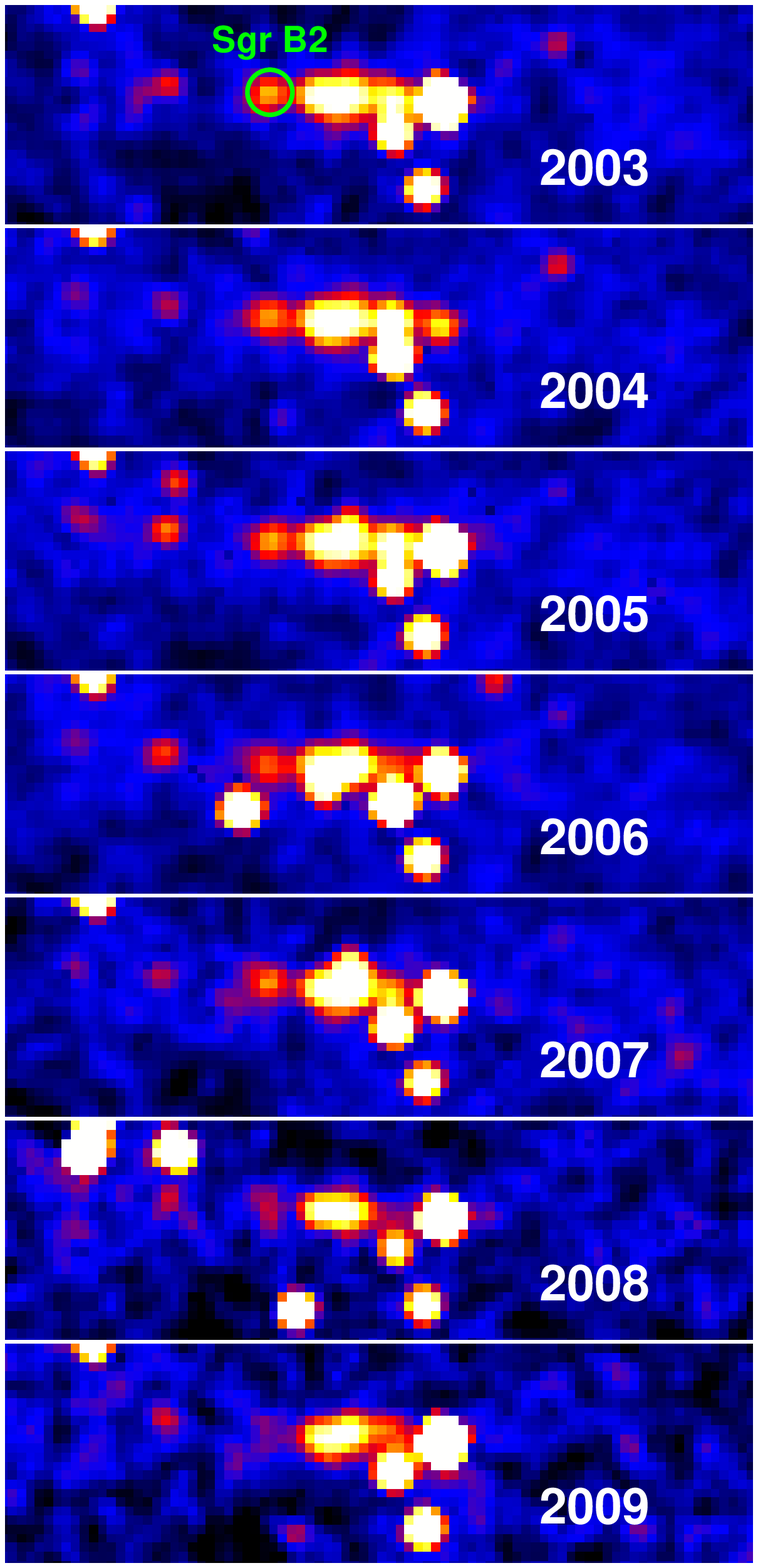}
\caption{The central degree as seen by IBIS/ISGRI on board INTEGRAL in the 20 to 60 keV energy range. Each row corresponds to a year from 2003 to 2009. The many sources in the GC display strong time dependence. The green circle shown in the top panel corresponds to the position of the core region of the Sgr B2 cloud -see the text. The associated hard X-ray source, IGR J17475-2822, shows a clear decline in flux after the 7 years of monitoring.\label{fig:maps}}
\end{figure*}

\subsection{IBIS/ISGRI data}

Since its launch in October 2002, INTEGRAL \citep{Winkler03} has been performing deep observations of the Galactic Centre region for more than 1 Ms every 6 months, first as part of the core program Galactic Centre Deep Exposure and then with the Galactic Centre Key Programme.  A number of dedicated pointings as well as Target of Opportunity  observations also add exposure to the GC thanks to the wide field of view of the main instruments on board INTEGRAL.  
We have therefore used all public data available within 13$^\mathrm{o}$ of the GC source position, as well as the latest observations of the Galactic Centre Key Progamme. We applied a number of selection criteria on the science windows' (i.e. individual pointings) quality so as to improve the overall dataset quality (imposing an exposure larger than 600s per pointing and rejecting images with bad significance distribution). This yields a number of 7600 science windows for a total of about 20 Ms of exposure. 

The size of the point spread function (PSF) is of fundamental importance when determining the contribution of the various sources in the GC. We therefore focused only on the IBIS/ISGRI data with its 12' PSF \citep{Lebrun03}. We analyzed the whole dataset using the Off-line Science Analysis (OSA) software version 7.0 with an improved energy reconstruction taking into account gain losses during the whole mission. We produced images of all science windows in 3 energy bands: 20-40 keV, 40-60 keV and 60-100 keV. Individual images were then mosaicked. A close-up of the Galactic central degrees is visible in Fig \ref{fig:IsgriImage}.  In order to extract regularly sampled lightcurves, we divided the observations into 14 periods (2 per year), each corresponding to an observation window lasting up to one month around September and March of each year, and we produced the 14 images of the Galactic Centre spanning from March 2003 up to September 2009.

Imperfect background and detector uniformity corrections introduce systematics biases and errors in the final mosaic images. The large number of bright sources in the Galactic central region also creates important systematic effects because of the remaining uncertainties in the IBIS/ISGRI PSF model that is subtracted from each individual image to remove ghost sidelobes \citep{Goldwurm03}.  To take into account all these systematics in the final mosaics, we applied a correction factor to the variance skymaps  to recover a normalized significance distribution of each image. 

To extract count rates of sources in the central degree where source confusion is large, we have to fit simultaneously the fluxes of all of the 11 sources present there. Since Sgr B2 is in a somewhat less crowded region, it is sufficient in this case to model the fluxes of just three point sources (namely 1E 1743.1-2843, IGR J17497-2821 and IGR J17475-2822/Sgr B2) using a gaussian PSF of width $\sigma$=0.1$^o$ in a 1$^\circ$ wide square region around the position of Sgr B2.

Since there is a correlation from bin to bin, due to the correlation technique used to make images with coded mask instruments, we cannot use the errors computed with the chi-square fit for the source fluxes. We therefore used the corrected image variance as the error on the source fluxes, which is a conservative estimate.

The ISGRI camera suffers from efficiency losses with time. To take these into account, we performed the same analysis on the calibration observations of the Crab Nebula taken on-axis during each of the 14 observation windows on the GC. We got a lightcurve of the measured Crab count rate in each energy band showing a clear decrease (about 5\% during the whole mission in the 20-40 keV energy range),  which is well fitted by a linear  decay. This provided us a  time-dependent correction to the measured count rates which we applied to the GC sources. As a verification procedure, we used  the Ophiuchus cluster as a secondary calibrator. It is visible in all GC mosaics and is  extended with respect to the  ISGRI PSF \citep{Eckert08}. We extracted its flux in each of the 14 mosaics and found the resulting lightcurve, after application of the flux correction,  to be compatible with a constant.

\subsection{XMM-Newton data}

Sgr B2 was observed  by XMM-Newton for 50 ks in September 2004 (ObsId: 0203930101). Event files were cleaned for flaring events and background-subtracted mosaicked images of the EPIC/MOS instruments have been produced. The background used for the subtraction is taken from reference high-latitude observations obtained in the same observing mode and filter and includes instrumental as well as cosmic backgrounds \citep{Carter07}. Prior to its subtraction, the background was normalized using the measured flux in the 10-12 keV energy range. To correct for the exposure and the vignetting, the mosaicked background-subtracted count images were divided by their mosaicked exposure maps which include the vignetting corrections.  The adaptative smoothing tool \texttt{asmooth} was applied to the images.

Images were produced in the 4-6 keV and in the 6.3 - 6.48 keV energy ranges. To remove the continuum from the 6.4 keV line image, we used the  background subtracted 4-6 keV image, renormalized according to the  fitted shape of the overall spectrum in the field of view.

We extracted spectra only from the 2004 observation, which was taken after the INTEGRAL launch and can be used for combined spectral fitting. We used a 4.5$^{\prime}$ radius extraction region. Because of the large spectral extraction region radius, we extracted background events from empty field observations taken in a similar time frame\citep{Carter07}.   

\section{Results}

\subsection{Time-varying hard X-ray emission from Sgr B2}

In the complete mosaic (covering 7 years of data), the source IGR J17475-2822,  associated with Sgr B2, is detected at a confidence level of more than 50 $\sigma$; see Fig \ref{fig:IsgriImage}. We left its position free in the multi-source fitting procedure described in section 2.1 and found it to be l = 0.67$^\mathrm{o}$, b = -0.05$^\mathrm{o}$ with a 90\% confidence radius of 0.7' \citep{Gros03}, in perfect agreement with the Sgr B2 core (M0.66--0.02) position, see the comparison with the 2004 XMM-Newton image in the Fe K$\alpha$ band (figure \ref{fig:IsgriPos}). Nevertheless, the large IBIS PSF encloses a significant part of the Sgr B2 complex, so that a contribution from neighbouring structures such as M0.74--0.09 cannot be excluded.  In particular, the larger and more diffuse component of the molecular cloud should be contributing to the hard X-ray flux. But its flux should be diluted because of the extent relative to the PSF size. If we assume that the envelope diameter is of the order of 45 pc \citep{Goldsmith90}, the expected diameter at 8 kpc is 20$^{\prime}$, significantly larger than the IBIS PSF. This implies a reconstructed peak flux of about 40\% of its equivalent point source flux \citep{Renaud06}.

Using the 2004 XMM-Newton/EPIC observation of the region, we checked that no point source in the field is bright enough to produce such a strong emission, so that we can unambiguously associate the hard X-ray source with diffuse emission from the Sgr B2 cloud. We also fitted the INTEGRAL source position at different time periods during the 7 years of observation but found no significant displacement of the source position.

\begin{figure*}
\epsscale{1.}
\plotone{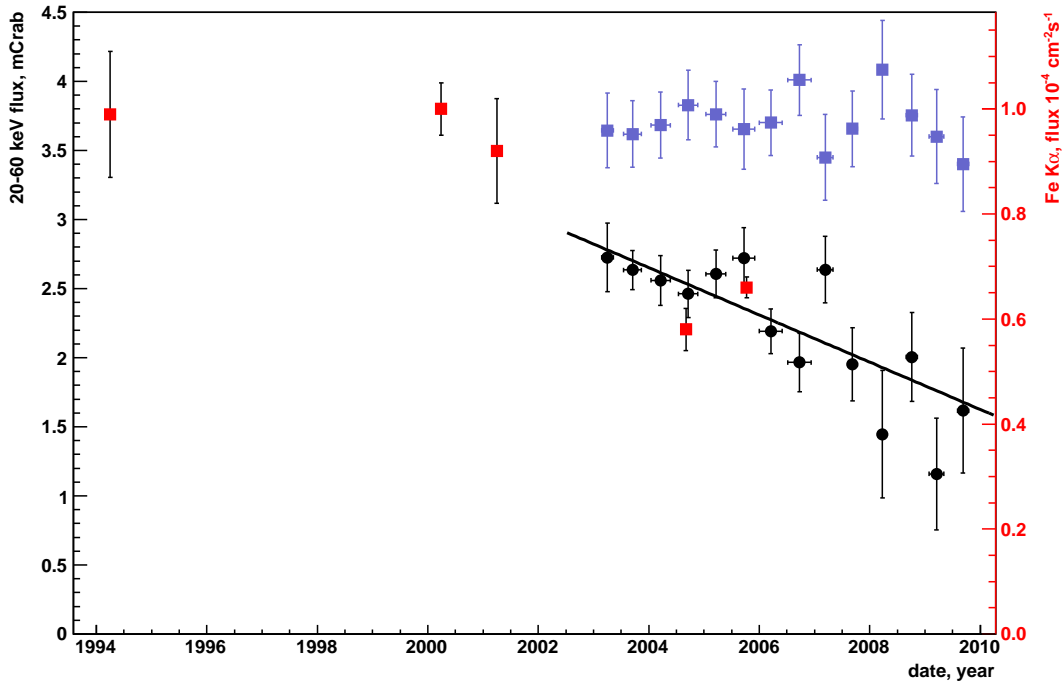}
\caption{{\it (Black circles)} Light curve of the corrected 20-60 keV flux in mcrab of Sgr B2 as measured with IBIS/ISGRI from 2003 to 2009. 
The light curve of the secondary calibrator (Ophiuchus cluster) obtained in the same conditions is shown for comparison ({\it Light blue squares}).
Superimposed ({\it red squares}) are the Fe K$\alpha$ line fluxes obtained by \citet{Inui09} with ASCA, Chandra, XMM-Newton and Suzaku observations taken from 1994 to 2006. The hard X-ray flux shows a clear decay of up to 40\% of the measured Sgr B2 flux from 2003 to 2009. A chi-square hypothesis test favors a linearly decreasing flux hypothesis over a constant flux at a 4.8$\sigma$ level, the line shows the best fit. The 6.4 keV measurements show a similar trend after the year 2000, the flux being apparently constant from 1994 to 2000.\label{fig:LC}}
\end{figure*}

Looking at INTEGRAL images of the GC we produced on a year timescale, most of the sources in the Galactic central degrees show strong variability on shorter time scales, as was noted by \citet{Kuulkers07}; see Fig \ref{fig:maps}. One can also note a clear but slow decrease of the flux of the source compatible with Sgr B2. To properly quantify this effect, we built a lightcurve of the Sgr B2 flux over the 7 years of INTEGRAL observations. To do so, we fixed the IGR J17475-2822 source position to the one obtained  in section 3.1 and extracted its count-rate in each of the 14 time periods using the multi-source fitting procedure. To correct for the varying instrument response, we applied the count-rate correction described in section 2.1. The resulting lightcurve is shown on figure \ref{fig:LC}. It is the most regularly sampled lightcurve of the non-thermal emission from this molecular cloud obtained so far, with a data point every six months. It is also the only one in the hard X-ray domain. We compare it to the 6.4 keV measurements of \citet{Inui09}. 

The Sgr B2 20-60 keV emission shows indications of a count rate decrease since 2003, reaching  40\% in 2009 compared to the 2003 flux.
 To estimate the significance of the effect, we performed an F-test comparing the $\chi^2$ of a constant flux fit ($\chi^2= 40.06$ d.o.f. = 13) and a linearly varying one ($\chi^2= 15.16$ d.o.f. = 12). The varying flux is preferred at the 4.8$\sigma$ level over a flat light curve.  The decay time  (time to divide the flux by two) is found to be $\tau = \left(4.0\ \pm\ 0.8\right)\times 10^3 $ days, that is, 8.2 $\pm$ 1.7 yrs, which is compatible with the light crossing time of the molecular cloud core and is compatible with previous expectations \citep{Sunyaev98}. The amplitude of the variation is consistent with that found by \citet{Inui09}.

\subsection{Broadband spectral analysis}

Sgr B2 is a faint object located in a crowded region containing more than a hundred sources within the IBIS field. Therefore its spectrum cannot be extracted using standard OSA spectral routines. We have built mosaics in 15 energy bands ranging from 18 to 300 keV, grouping the data in 2 year intervals.   We used a fitting procedure similar to that described in section 2.1 to fit the flux of Sgr B2 as well as that of neighboring sources and the background in each energy band. The position of Sgr B2 was fixed to the position fitted in the complete dataset in the 20-60 keV energy range.  We applied the same procedure to estimate the errors on the flux measurement to take into account systematic errors due to imperfect background and source side lobe subtraction. 

Because of the evolution of the instrument response with time, we computed mean ancillary response functions, {\it arf},  for each of the 2-year periods using matrices distributed with the OSA.  Applying these to the Crab Nebula observations, we estimated a systematic uncertainty of $\sim$ 8\%, which we applied to all spectral fits of the IBIS/ISGRI data.

We performed a broadband fit to the 2004 XMM-Newton EPIC/MOS and PN and the 2003-2004 IBIS/ISGRI spectra. We followed an approach similar to that of \citet{Ryu09} but considered only events above 3.5 keV to avoid modeling the low energy background in the GC and did not take into account the Cosmic X-ray Background (CXB) contribution which is subtracted by the blank sky events.  We therefore used  an \texttt{APEC} model for the hot thermal emission (the so-called 8 keV plasma) and we modeled the spectrum from the molecular cloud as the sum of a power law and of two gaussians for the iron K$\alpha$ and K$\beta$ lines (respectively at 6.4 and 7.06 keV). All the emission components were subject to line of sight absorption due to a foreground interstellar column density of $\mathrm{\sim 6\times 10^{22}\ cm^{-2}}$ \citep{Sakano02}.  The resulting XSPEC model is therefore $\mathrm{wabs} \times\mathrm{ ( apec + gaus + gaus + wabs} \times \mathrm{ pegpw ) }$.
The energies of the two iron lines are fixed to 6.4 and 7.06 keV respectively, and we impose the ratio of Fe K$\beta$ to K$\alpha$ to be equal to 15\% \citep{Murakami00}. We obtain a satisfactory fit  with a reduced $\chi^2$ of 1.19.  The plasma temperature is 5.8 $\pm$ 0.3 keV and the power-law component has a spectral index of 2.07 $\pm$ 0.05 and a 2-100 keV unabsorbed flux normalization of 8.4 $\pm$ 0.4 $\mathrm{\times 10^{-11} erg\ cm^{-2}\ s^{-1}}$. The flux of the iron K$\alpha$ line is found to be 8.9 $\pm$ 0.6  $\mathrm{\times 10^{-5}\ cm^{-2}\ s^{-1}}$  . The intrinsic absorption is found to be 6.8 $\pm$ 0.5 $\mathrm{\times 10^{23}\ cm^{-2}}$, which is very large but still lower than  estimates of the maximum column density of the cloud \citep{Protheroe08}. The foreground absorption is 6.9 $\pm$ 0.5 $\mathrm{\times 10^{22}\ cm^{-2}}$, consistent with other estimates \citep{Ryu09}.

In the following, we try to apply detailed emission models to the Sgr B2 spectrum to constrain the origin of the emission.

\section{On the nature of the hard X-ray emission}

Our observation of a clear decay of the hard X-ray flux is consistent with that reported earlier for the 6.4 keV line emission by \citet{Inui09} -- see Fig \ref{fig:LC}. We measure a timescale of 
8.2 $\pm$ 1.7 yrs for a 40\% linear decrease of the hard flux compared to what was measured in 2003. The observation of such a rapidly varying flux is very difficult to explain with cosmic ray particles. \citet{Dogiel09} have invoked a population of 100 MeV protons as a possible origin of the hard X-ray emission from the GC clouds. But since the relevant  Coulomb cooling time, even in a dense cloud such as Sgr B2, is of the order of several hundred years, we can exclude the possibility that such a component makes a significant contribution to the observed X-ray emission. 

High energy ($\sim$ GeV) electrons can also produce hard X-ray emission through inverse Compton scattering on the ambient IR emission which is expected to be high in the Sgr B2 core because of its huge luminosity. \citet{Goicoechea04} found a far IR luminosity of $\mathrm{8.5 \times 10^6 \ L_{\odot}}$ originating mainly from the Sgr B2(M) and (N) cores. In a radius of 1 pc, this creates a huge radiation density of  $\mathrm{6 \times 10^3 eV\ cm^{-3}}$ in FIR photons. \citet{Leahy93} show that for a $1/r^2$ density profile, the Fe K$\alpha$ EW can reach about 1 keV for an optical depth of the order of 1. With column densities towards the core of the cloud reaching $10^{24}\ \mathrm{cm^{-2}}$ \citep{Protheroe08}, inverse Compton emission of high energy electrons in the centre of Sgr B2 might create a signal similar to that which we observe.  Nevertheless we are faced with similar problems: with densities reaching  $\mathrm{10^5\ cm^{-3}}$ the dominant energy loss mechanism is bremsstrahlung, but its characteristic timescale is of the order of 50 yrs.   

We are left only with low energy cosmic ray electron models. In  this case, variability of the overall flux is not excluded since the energy loss timescale for ionization losses of $\sim$ 100 keV electrons is of the order of or less than 1 yr in dense environments of $\mathrm{10^{5}\ cm^{-3}}$. We discuss  this possibility in more detail in the next subsection using a detailed spectral model of the emission.

\subsection{A spectral model of LECRE emission}

In order to test for a possible LECRE origin of the 6.4 keV line and hard X-ray continuum emission of Sgr B2, we implemented a model of low energy cosmic ray electron interactions in dense matter to predict the X-ray spectrum of their emission and compare it to the Sgr B2 measured spectrum. We used a steady-state, thick target model, in which accelerated electrons with given  energy spectra are injected at a constant rate into the interaction region and  produce atomic reactions as they slow down to energies below the thresholds of  the various reactions \citep{Tatischeff98}. It includes the non-thermal X-ray  emission from the secondary, knock-on electrons, which mainly result from  ionization of ambient $\mathrm{H_2}$ molecules and He atoms. The required differential  ionization cross sections are calculated from the relativistic binary  encounter dipole theory \citep{Kim00}. The  continuum emission is due to bremsstrahlung of both primary and secondary  electrons. Calculations of the bremsstrahlung spectra are based on \citet{Strong00}. The line emission results from the  filling of inner-shell vacancies produced by both primary and secondary  electrons in ambient atoms. We considered the K$\alpha$ lines of C, N, O, Ne and  Mg (although they are strongly absorbed at the GC) and the K$\alpha$ and K$\beta$ lines from Si, S, Ar, Ca, Fe and Ni. The electron-induced  K-shell ionization cross sections are from Quarles semi-empirical formula \citep{Quarles76}

The energy spectrum of the fast electrons injected in the X-ray production  region is a power-law, $dQ/dt(E) \propto E^{-s}$ between $\mathrm{E_{min}}$ and $\mathrm{E_{max}}$.  The minimum energy $\mathrm{E_{min}}$ is set to 1 keV. The spectral index $s$ and the maximum  energy $\mathrm{E_{max}}$ are free parameters. So is the ambient medium abundance. The model has therefore 3 free parameters. 

We note that photoelectric absorption of bremsstrahlung X-rays in the source also  contributes to fluorescence line emission. The equivalent width of the fluorescence Fe K$\alpha$ line  can be estimated from the  absorbing hydrogen column density of the source, $\mathrm{N_H(source)}$, to be
$\mathrm{EW = 0.07 \times abund \times (N_H(source) / 10^{23} cm^{-2})}$ keV. This approximation is valid for $\mathrm{N_H(source) < 10^{24} cm{^{-2}}}$ \citep{Leahy93}. 

The model has been implemented as an XSPEC table model and tested on our broadband spectrum of Sgr B2.
We used the following model: $\mathrm{wabs} \times\mathrm{ ( apec + wabs} \times \mathrm{lecre ) }$. To limit the number of free parameters, we fixed the temperature of the plasma at 6.5 keV and its metallicity to solar abundances (see \citealt{Koyama07,Ryu09}). 
We obtain a satisfactory fit to the XMM-Newton  and ISGRI data with a reduced $\chi^2$ of 1.16 (1383 d.o.f.).  The CR electron spectral index is not well constrained  but it is found to be hard, with $s \sim 1.5$, compared to typical acceleration processes. The metallicity of the molecular cloud relative to solar  (as given by \citealt{Lodders03}) is $A = 3.1 \pm 0.2$ which is significantly less than that estimated by \citet{Revnivtsev04}, but still too large compared to current measurements. Thanks to the spectral coverage, the fit provides a constraint on the maximum energy of the electrons such that $E_{max} = 200 \pm 50$ keV. This shows that a single population cannot be at the origin of the hard X-ray and TeV emissions, as has already been by \citet{Crocker07}.  The CR luminosity  required to sustain the observed flux is $ 8.9 \pm 1.3\ \times 10^{39}$ erg/s. This luminosity requirement is slightly lower than that estimated by \citet{Revnivtsev04} but is still $\sim$ 30\% of the total far infrared luminosity of Sgr B2 \citep{Goicoechea04}. We stress that since most of this power is dissipated by electrons of energy close to $E_{max}$ the hard X-ray spectrum is important to constrain this parameter.  The molecular cloud absorption column density is found to be 5.8 $\pm$ 0.3 $\mathrm{\times 10^{23}\ cm^{-2}}$. The maximal column density estimated by \citet{Protheroe08} is  12 $\mathrm{\times 10^{23}\ cm^{-2}}$, suggesting that the cosmic ray production and interaction would have to take place deep inside Sgr B2.

\citet{Giveon02} have measured the GC interstellar medium metallicity to be twice solar. Similarly, studies of several red giant and supergiant stars in the GC have shown that their metallicity is only slightly above solar \citep{Cunha08,Davies09}. The metallicity required to reproduce the iron line equivalent width is therefore too large.

 Besides, the huge power input is a strong limitation, since it is close to the total bolometric luminosity of the cloud.  The decay time is also a strong limitation of this scenario: the size of the 6.4 keV line emission region associated with the hard X-ray continuum we measure implies that either the subrelativisitic electron production is distributed inside the cloud or that a huge number of them have been produced in a limited region and diffuse into the cloud. Both solutions are ruled out: the first one because we should not expect correlated variation of the electron sources all over the cloud, and the second one because the time required for electrons to diffuse in a region as large as that of the diffuse 6.4 keV line  emission is much larger than the measured decay time of the signal. We can therefore exclude LECRE emission as the dominant contribution of the iron line and hard X-ray emission in Sgr B2.

\subsection{Hard X-ray reflection from an external distant source}

Given the difficulties outlined above with cosmic-ray models, it is necessary to turn to the alternative interpretation in terms of X-ray reflection, for which variability on relatively short timescales is expected \citep{Sunyaev98}. When the illuminating source is switched off, the decay of the scattered flux should occur on a timescale of the light crossing time for the cloud.  The actual location of the illuminating source has then to be determined. An internal source might also create such a decrease if it stopped emitting in the recent past.  Nevertheless, \citet{Revnivtsev04} reject this possibility due to the relative stability of the measured Fe K$\alpha$ flux between 1993 and 2000, which would imply that the source was still active in the last decade. Since the column density can reach $\mathrm{10^{24}\ cm^{-2}}$, we might expect an opacity of about 0.66 -- not large enough to prevent the direct detection of the source during the various  observations of the molecular cloud in this period.  It is therefore likely that the source is external.

We developed an XRN model assuming a relatively simple geometry for the cloud. Following the approach of \citet{Murakami00}, we use a 45 pc diameter cylindrical cloud with a symmetry axis parallel to the line of sight. The primary irradiating X-ray source is supposed to be located at the side of the cloud, at  an angle of 90$^o$ with respect to the line of sight. The region enclosing the cloud is divided for the numerical simulations into 225$\times$225$\times$113 cubic cells of 0.2 pc on a side. 

 The density distribution is assumed to be that of \citet{Goldsmith90}, with a very high density 5-pc-wide core surrounded by a more diffuse envelope.  The total  mass of the cloud is normalized to be $\mathrm{1.9\ 10^6\ M_{\odot}}$, which is the virial mass of Sgr B2 as  recently estimated by \citet{Protheroe08}.

The probability of scattering of a primary X-ray photon is calculated from  both the coherent (Rayleigh) and incoherent (Compton) scattering cross sections, taking into account the 13 most abundant elements in the ambient medium  (H, He, C, N, O, Ne, Mg, Si, S, Ar, Ca, Fe and Ni). These cross sections are taken from  the XCOM photon cross sections database \citep{XCOM}. Above a few keV depending on the composition, the total scattering cross section is dominated by the incoherent process. The energy of the scattered photon is then given by  $E_X^{\prime} = E_X m_e c^2 / (E_X+m_ec^2)$ ($E_X$ being the energy of the primary photon and $m_e$  the electron mass), which is appropriate for a Compton scattering at 90$^o$.

The fluorescent K$\alpha$ and K$\beta$ lines of Si, S, Ar, Ca, Cr, Mn, Fe and Ni are considered. We also include the K$\alpha$  line of C, N, O, Ne and Mg although these lines are strongly absorbed in the present case. The K-shell photo-ionization cross  sections are from \citet{Verner95}. 
Photoelectric absorption of both the primary and secondary X-rays is calculated from the elemental cross sections of \citet{Balucinska92}. It thus depends on the cloud composition.

The resulting spectra were then implemented in an XSPEC table model with 3 parameters: the irradiating source spectral index (which can vary from 1 to 5), the ambient medium abundance, which can vary from 0.5 to 3 times  solar (the solar metallicity is taken from \citet{Lodders03})  and the flux normalization.

\begin{figure*}
\epsscale{1.2}
\begin{center}
\includegraphics[scale=0.55,angle=-90]{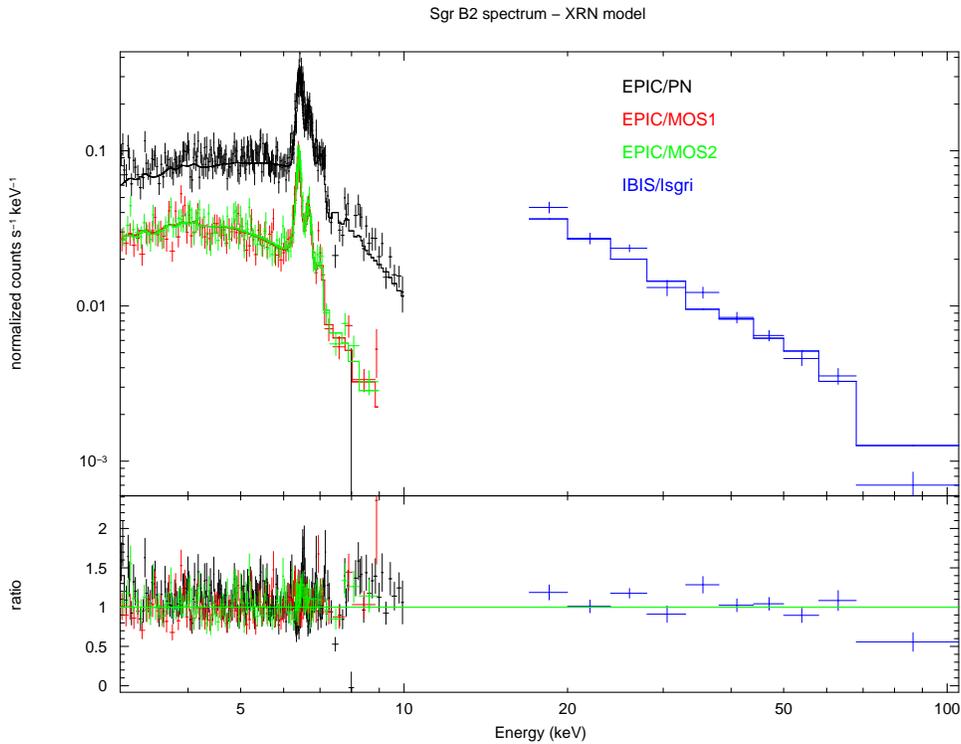}
\end{center}
\caption{Broadband spectrum of the Sgr B2 molecular cloud fitted with a X-ray nebula spectral model (see text for details)\label{fig:specxrn}}
\end{figure*}

We used a total model similar to that used in the previous section: $\mathrm{wabs} \times\mathrm{ ( apec + xrn ) }$.
We obtain a satisfactory fit to the XMM-Newton (2004) and ISGRI data (2003-2004) with a reduced $\chi^2$ of 1.16 (1384 d.o.f.).  The illuminating source power law spectral index is found to be $\Gamma \sim 2$, the abundance relative to solar $A = 1.3 \pm 0.1$ and the average luminosity of the illuminating source required to sustain the observed flux is $\mathrm{L(1-100\ keV) = 1.1 \times 10^{39} \left(d/100pc\right)^2}$ erg/s, where d is the distance of the source to Sgr B2.

As has been noted in previous work, if the illuminating source were to be at a distance of $\sim$ 10 pc its luminosity would have been larger than $10^{37}$ erg/s for several years and it should have been detected in the past decades, but if the source is farther away the luminosity constraints leave only Sgr A$^{\star}$ as a viable candidate \citep{Koyama96, Murakami00,Revnivtsev04}. In the latter case, Sgr A$^*$ should also be responsible for the measured Fe K$\alpha$ in some other molecular clouds of the CMZ. The observations of time dependent effects in other places is therefore very strong evidence of the XRN nature of the Sgr B2 hard X-ray emission and supports Sgr A$^*$ as the origin of the illuminating flux \citep{Ponti10}.
  
In a  recent analysis of VLBI observations, Reid et al. (2009) have measured the parallax of the Sgr B2 (M) and Sgr B2 (N) dust condensates. The authors have estimated the Galactocentric radius at $R_0 = 7.9^{+0.8}_{-0.7}$ kpc and the proper motion of Sgr B2 relative to Sgr A$^{\star}$ suggests that the former is $130 \pm 60$ pc closer to us than the latter. This fact is also suggested by the study of photoelectric absorption in front of the various molecular clouds inside the Sgr B complex \citep{Ryu09}. In a scenario where intense Sgr A$^{\star}$  activity is  the origin of the hard X-rays from the cloud one can date the end of the high state period to be $100^{+55}_{-25}$ years ago, notably different from the widely discussed 300 years obtained assuming the GMC is at the same distance as Sgr A$^*$ (see e.g. \citet{Koyama96}).

 This geometry is inconsistent with the model we use which requires orthogonal photon incidence on the cloud. At 130 pc in front of Sgr A$^{\star}$, several effects should modify the resulting spectrum. First, the mean column density seen by scattered photons is larger, increasing the absorption with respect to the orthogonal scenario. Second, the Compton diffusion cross section is larger for  scattering angles smaller than 90$^{\circ}$ at high energies. The resulting spectrum should therefore be harder and brighter in the hard X-ray domain and more attenuated under $\lesssim 15$ keV.  To estimate the effect on the measured spectral parameters, we have performed simple Monte-Carlo simulations of photon propagation in a spherical uniform cloud. For clouds mass comparable to that of Sgr B2, we found out that the differences on the measured spectral index could be up to 0.2 in the 15-100 keV range for $\cos\theta$ ranging from 0 to 0.9 (corresponding to a cloud 190 pc in front of the GC). The total scattered flux also increases by up to 30\% in the 5 to 100 keV band. 
Therefore, taking into account uncertainties due to the relative position of SgrB2 and Sgr A$^{\star}$, the actual luminosity from the flare should range between 1.5 and 5 $10^{39}$ erg/s and its spectral index should be $\Gamma = 2 \pm 0.2$.

\section{Conclusions}

INTEGRAL has been regularly monitoring the Galactic Centre for 7 years. Using the complete dataset, we confirm the result from \citet{Revnivtsev04} that IGR J17475-2822 corresponds to hard X-ray ($>$20 keV) emission from the Sgr B2 molecular cloud. In particular, it is in good positional agreement with the very massive and dense core of the cloud. The larger and more diffuse component might be contributing but its flux should be diluted because of its extent relative to the PSF size. Since the mean density is of the order of $\mathrm{1-3 \times 10^3 cm^{-3}}$, the envelope column density is smaller by a factor of a few compared to the core column density (which is close to $\mathrm{10^{24} cm^{-2}}$), so that the diffuse hard X-ray contribution has to be much smaller than that of the core. 

The regular monitoring  shows that the Sgr B2 20-60 keV emission is fading, with a flux decrease of about 40\% after 7 years of observations. A test of hypothesis shows that a linear decay is favored compared to a constant flux at the 4.8 $\sigma$ confidence level.  The linear decay characteristic timescale is found to be $\sim$ 8 yrs,  comparable with the light-travel time across the cloud core.  This result is obtained thanks to the very deep exposure obtained by INTEGRAL on the GC regions and does not suffer from instrument to instrument  intercalibration errors. It is consistent with the results obtained comparing the 6.4 keV fluxes measured by Chandra, XMM-Newton and Suzaku from 2000 to 2005 \citep{Inui09}. 

Cosmic ray models of the iron line and continuum emission are put to a test by this fast variability. Inverse bremsstrahlung emission from subrelativistic ions is ruled out because of the long cooling timescale of these particles in the dense environment of the cloud core.
Low energy cosmic ray electrons are not formally ruled out: the lifetime of $\sim$ 100 keV electrons in the dense matter of Sgr B2 is smaller  than the measured decay time, so that if the cosmic-ray injection had ceased with a similar or smaller timescale, we might obtain a similar flux decrease.  Nevertheless, combining the measured decay with the large electron injection power required (which we fitted to be of the order of $10^{40}$ erg/s) we can exclude this scenario as well.

Sgr B2 is therefore most likely a reflection nebula, as was initially proposed by \citet{Sunyaev93} and \citet{Koyama96}.  The location of the illuminating source is yet to be determined unequivocally. As was already noted by several works, an internal or a nearby illuminating source should have been detected in the last decade or so. The most likely explanation is therefore a period of high state of Sgr A$^{\star}$. The observation of variability of the Fe K$\alpha$ line in other massive molecular clouds of the CMZ strongly supports this hypothesis \citep{Ponti10}, since several distinct bright illuminating sources active for at least several years would be required to light all these clouds. They could be illuminated by the same outburst from Sgr A$^*$. 

Using a model based on the approach of \citet{Murakami00}, we tested this scenario and found that the illuminating spectrum must be hard (photon index $\sim$ 2), which is well in the range of measured Low Luminosity Active Galactic Nuclei (LLAGN) spectral indices \citep{Gu09,Pellegrini00}. Since parallax measurements place Sgr B2 130 $\pm$ 60 pc in front of Sgr A$^{\star}$  \citep{Reid09}, our measurement implies a mean luminosity of 2-5 $\times 10^{39}$ erg/s, an increase of more than 5 orders of magnitude compared to the current one placing it among faint LLAGNs \citep{Ho08,Gu09}. At the end of the outburst the variability of Sgr A$^*$ is constrained to be faster than the measured decay rate, which is comparable to the long term variability observed in M81$^*$ \citep{Iyomoto01,LaParola04}. 
 The end of the episode of high luminosity in Sgr A*, which is currently being mirrored by Sgr B2, occurred about 100 yrs ago, with the intervening period seeing the output of the central source fall by some 5 orders of magnitude. Assuming that Sgr B2 is at the same distance that Sgr A$^{\star}$, \citet{Inui09} have proposed that the luminosity of the latter has been continuously decreasing with a constant decay time for 300 years. The new position implies that Sgr A$^{\star}$ luminosity had to decrease on a much shorter timescale. The origin of such large swings in the luminosity of the supermassive black hole at the centre of our Galaxy is presently unknown but the subject of intensive on-going research.

\acknowledgments
 This work is based on observations with INTEGRAL, an ESA project with instruments
    and science data centre funded by ESA member states , Czech
    Republic and Poland, and with the participation of Russia and the USA.

Based on observations with XMM-Newton, an ESA Science Mission with instruments and contributions directly funded by ESA Member States and the USA (NASA).

We thank M. Falanga for giving us access to the data from the ToO observation of IGR J17511-3057.

This work has been supported by ANR grant ANR-JCJC-06-0047.

{\it Facilities:} \facility{INTEGRAL (IBIS)}, \facility{XMM (EPIC)}.

\bibliographystyle{apj}                       

\bibliography{SgrB2}

\begin{thebibliography}{59}
\expandafter\ifx\csname natexlab\endcsname\relax\def\natexlab#1{#1}\fi

\bibitem[{{Aharonian} {et~al.}(2006){Aharonian}, {Akhperjanian}, {Bazer-Bachi},
  {Beilicke}, {Benbow}, {Berge}, {Bernl{\"o}hr}, {Boisson}, {Bolz}, {Borrel},
  {Braun}, {Breitling}, {Brown}, {Chadwick}, {Chounet}, {Cornils},
  {Costamante}, {Degrange}, {Dickinson}, {Djannati-Ata{\"i}}, {Drury}, {Dubus},
  {Emmanoulopoulos}, {Espigat}, {Feinstein}, {Fontaine}, {Fuchs}, {Funk},
  {Gallant}, {Giebels}, {Gillessen}, {Glicenstein}, {Goret}, {Hadjichristidis},
  {Hauser}, {Hauser}, {Heinzelmann}, {Henri}, {Hermann}, {Hinton}, {Hofmann},
  {Holleran}, {Horns}, {Jacholkowska}, {de Jager}, {Kh{\'e}lifi}, {Klages},
  {Komin}, {Konopelko}, {Latham}, {Le Gallou}, {Lemi{\`e}re},
  {Lemoine-Goumard}, {Leroy}, {Lohse}, {Marcowith}, {Martin},
  {Martineau-Huynh}, {Masterson}, {McComb}, {de Naurois}, {Nolan}, {Noutsos},
  {Orford}, {Osborne}, {Ouchrif}, {Panter}, {Pelletier}, {Pita},
  {P{\"u}hlhofer}, {Punch}, {Raubenheimer}, {Raue}, {Raux}, {Rayner}, {Reimer},
  {Reimer}, {Ripken}, {Rob}, {Rolland}, {Rowell}, {Sahakian}, {Saug{\'e}},
  {Schlenker}, {Schlickeiser}, {Schuster}, {Schwanke}, {Siewert}, {Sol},
  {Spangler}, {Steenkamp}, {Stegmann}, {Tavernet}, {Terrier}, {Th{\'e}oret},
  {Tluczykont}, {van Eldik}, {Vasileiadis}, {Venter}, {Vincent}, {V{\"o}lk}, \&
  {Wagner}}]{Aharonian06}
{Aharonian}, F., {et~al.} 2006, \nat, 439, 695

\bibitem[{{Balucinska-Church} \& {McCammon}(1992)}]{Balucinska92}
{Balucinska-Church}, M., \& {McCammon}, D. 1992, \apj, 400, 699

\bibitem[{{B{\'e}langer} {et~al.}(2006){B{\'e}langer}, {Goldwurm}, {Renaud},
  {Terrier}, {Melia}, {Lund}, {Paul}, {Skinner}, \& {Yusef-Zadeh}}]{Belanger06}
{B{\'e}langer}, G., {et~al.} 2006, \apj, 636, 275

\bibitem[{{Berger} {et~al.}(2009){Berger}, {Hubbell}, {Seltzer}, {Chang},
  {Coursey}, {Sukumar}, \& {Zucker}}]{XCOM}
{Berger}, M.~J., {Hubbell}, J.~H., {Seltzer}, S.~M., {Chang}, J., {Coursey},
  J.~S., {Sukumar}, R., \& {Zucker}, D. 2009, XCOM: Photon Cross Sections
  Database, \texttt{http://www.nist.gov/physlab/data/xcom/index.cfm}

\bibitem[{{Carter} \& {Read}(2007)}]{Carter07}
{Carter}, J.~A., \& {Read}, A.~M. 2007, \aap, 464, 1155

\bibitem[{{Crocker} {et~al.}(2007){Crocker}, {Jones}, {Protheroe}, {Ott},
  {Ekers}, {Melia}, {Stanev}, \& {Green}}]{Crocker07}
{Crocker}, R.~M., {Jones}, D., {Protheroe}, R.~J., {Ott}, J., {Ekers}, R.,
  {Melia}, F., {Stanev}, T., \& {Green}, A. 2007, \apj, 666, 934

\bibitem[{{Cunha} {et~al.}(2008){Cunha}, {Smith}, {Sellgren}, {Blum},
  {Ram{\'{\i}}rez}, \& {Terndrup}}]{Cunha08}
{Cunha}, K., {Smith}, V.~V., {Sellgren}, K., {Blum}, R.~D., {Ram{\'{\i}}rez},
  S.~V., \& {Terndrup}, D.~M. 2008, in IAU Symposium, Vol. 245, IAU Symposium,
  ed. {M.~Bureau, E.~Athanassoula, \& B.~Barbuy}, 339--342

\bibitem[{{Davies} {et~al.}(2009){Davies}, {Origlia}, {Kudritzki}, {Figer},
  {Rich}, \& {Najarro}}]{Davies09}
{Davies}, B., {Origlia}, L., {Kudritzki}, R., {Figer}, D.~F., {Rich}, R.~M., \&
  {Najarro}, F. 2009, \apj, 694, 46

\bibitem[{{de Cesare} {et~al.}(2007){de Cesare}, {Bazzano}, {Mart{\'{\i}}nez
  N{\'u}{\~n}ez}, {Stratta}, {Tarana}, {Del Santo}, \& {Ubertini}}]{deCesare07}
{de Cesare}, G., {Bazzano}, A., {Mart{\'{\i}}nez N{\'u}{\~n}ez}, S., {Stratta},
  G., {Tarana}, A., {Del Santo}, M., \& {Ubertini}, P. 2007, \mnras, 380, 615

\bibitem[{{Dogiel} {et~al.}(2009){Dogiel}, {Cheng}, {Chernyshov}, {Bamba},
  {Ichimura}, {Inoue}, {Ko}, {Kokubun}, {Maeda}, {Mitsuda}, \&
  {Yamasaki}}]{Dogiel09}
{Dogiel}, V., {et~al.} 2009, \pasj, 61, 901

\bibitem[{{Eckert} {et~al.}(2008){Eckert}, {Produit}, {Paltani}, {Neronov}, \&
  {Courvoisier}}]{Eckert08}
{Eckert}, D., {Produit}, N., {Paltani}, S., {Neronov}, A., \& {Courvoisier}, T.
  2008, \aap, 479, 27

\bibitem[{{Fryer} {et~al.}(2006){Fryer}, {Rockefeller}, {Hungerford}, \&
  {Melia}}]{Fryer06}
{Fryer}, C.~L., {Rockefeller}, G., {Hungerford}, A., \& {Melia}, F. 2006, \apj,
  638, 786

\bibitem[{{Ghez} {et~al.}(2008){Ghez}, {Salim}, {Weinberg}, {Lu}, {Do}, {Dunn},
  {Matthews}, {Morris}, {Yelda}, {Becklin}, {Kremenek}, {Milosavljevic}, \&
  {Naiman}}]{Ghez08}
{Ghez}, A.~M., {et~al.} 2008, \apj, 689, 1044

\bibitem[{{Gillessen} {et~al.}(2009){Gillessen}, {Eisenhauer}, {Fritz},
  {Bartko}, {Dodds-Eden}, {Pfuhl}, {Ott}, \& {Genzel}}]{Gillessen09}
{Gillessen}, S., {Eisenhauer}, F., {Fritz}, T.~K., {Bartko}, H., {Dodds-Eden},
  K., {Pfuhl}, O., {Ott}, T., \& {Genzel}, R. 2009, \apjl, 707, L114

\bibitem[{{Giveon} {et~al.}(2002){Giveon}, {Sternberg}, {Lutz}, {Feuchtgruber},
  \& {Pauldrach}}]{Giveon02}
{Giveon}, U., {Sternberg}, A., {Lutz}, D., {Feuchtgruber}, H., \& {Pauldrach},
  A.~W.~A. 2002, \apj, 566, 880

\bibitem[{{Goicoechea} {et~al.}(2004){Goicoechea},
  {Rodr{\'{\i}}guez-Fern{\'a}ndez}, \& {Cernicharo}}]{Goicoechea04}
{Goicoechea}, J.~R., {Rodr{\'{\i}}guez-Fern{\'a}ndez}, N.~J., \& {Cernicharo},
  J. 2004, \apj, 600, 214

\bibitem[{{Goldwurm} {et~al.}(2003){Goldwurm}, {David}, {Foschini}, {Gros},
  {Laurent}, {Sauvageon}, {Bird}, {Lerusse}, \& {Produit}}]{Goldwurm03}
{Goldwurm}, A., {et~al.} 2003, \aap, 411, L223

\bibitem[{{Gros} {et~al.}(2003){Gros}, {Goldwurm}, {Cadolle-Bel}, {Goldoni},
  {Rodriguez}, {Foschini}, {Del Santo}, \& {Blay}}]{Gros03}
{Gros}, A., {Goldwurm}, A., {Cadolle-Bel}, M., {Goldoni}, P., {Rodriguez}, J.,
  {Foschini}, L., {Del Santo}, M., \& {Blay}, P. 2003, \aap, 411, L179

\bibitem[{{Gu} \& {Cao}(2009)}]{Gu09}
{Gu}, M., \& {Cao}, X. 2009, \mnras, 399, 349

\bibitem[{{Ho}(2008)}]{Ho08}
{Ho}, L.~C. 2008, \araa, 46, 475

\bibitem[{{Inui} {et~al.}(2009){Inui}, {Koyama}, {Matsumoto}, \&
  {Tsuru}}]{Inui09}
{Inui}, T., {Koyama}, K., {Matsumoto}, H., \& {Tsuru}, T.~G. 2009, \pasj, 61,
  241

\bibitem[{{Iyomoto} \& {Makishima}(2001)}]{Iyomoto01}
{Iyomoto}, N., \& {Makishima}, K. 2001, \mnras, 321, 767

\bibitem[{{Kim} {et~al.}(2000){Kim}, {Santos}, \& {Parente}}]{Kim00}
{Kim}, Y., {Santos}, J.~P., \& {Parente}, F. 2000, \pra, 62, 052710

\bibitem[{{Koyama} {et~al.}(1996){Koyama}, {Maeda}, {Sonobe}, {Takeshima},
  {Tanaka}, \& {Yamauchi}}]{Koyama96}
{Koyama}, K., {Maeda}, Y., {Sonobe}, T., {Takeshima}, T., {Tanaka}, Y., \&
  {Yamauchi}, S. 1996, \pasj, 48, 249

\bibitem[{{Koyama} {et~al.}(2007){Koyama}, {Hyodo}, {Inui}, {Nakajima},
  {Matsumoto}, {Tsuru}, {Takahashi}, {Maeda}, {Yamazaki}, {Murakami},
  {Yamauchi}, {Tsuboi}, {Senda}, {Kataoka}, {Takahashi}, {Holt}, \&
  {Brown}}]{Koyama07}
{Koyama}, K., {et~al.} 2007, \pasj, 59, 245

\bibitem[{{Kuulkers} {et~al.}(2007){Kuulkers}, {Shaw}, {Paizis}, {Chenevez},
  {Brandt}, {Courvoisier}, {Domingo}, {Ebisawa}, {Kretschmar}, {Markwardt},
  {Mowlavi}, {Oosterbroek}, {Orr}, {R{\'{\i}}squez}, {Sanchez-Fernandez}, \&
  {Wijnands}}]{Kuulkers07}
{Kuulkers}, E., {et~al.} 2007, \aap, 466, 595

\bibitem[{{La Parola} {et~al.}(2004){La Parola}, {Fabbiano}, {Elvis},
  {Nicastro}, {Kim}, \& {Peres}}]{LaParola04}
{La Parola}, V., {Fabbiano}, G., {Elvis}, M., {Nicastro}, F., {Kim}, D.~W., \&
  {Peres}, G. 2004, \apj, 601, 831

\bibitem[{{LaRosa} {et~al.}(2005){LaRosa}, {Brogan}, {Shore}, {Lazio},
  {Kassim}, \& {Nord}}]{LaRosa05}
{LaRosa}, T.~N., {Brogan}, C.~L., {Shore}, S.~N., {Lazio}, T.~J., {Kassim},
  N.~E., \& {Nord}, M.~E. 2005, \apjl, 626, L23

\bibitem[{{Leahy} \& {Creighton}(1993)}]{Leahy93}
{Leahy}, D.~A., \& {Creighton}, J. 1993, \mnras, 263, 314

\bibitem[{{Lebrun} {et~al.}(2003){Lebrun}, {Leray}, {Lavocat}, {Cr{\' e}tolle},
  {Arqu{\` e}s}, {Blondel}, {Bonnin}, {Bou{\` e}re}, {Cara}, {Chaleil}, {Daly},
  {Desages}, {Dzitko}, {Horeau}, {Laurent}, {Limousin}, {Mathy}, {Mauguen},
  {Meignier}, {Molini{\' e}}, {Poindron}, {Rouger}, {Sauvageon}, \&
  {Tourrette}}]{Lebrun03}
{Lebrun}, F., {et~al.} 2003, \aap, 411, L141

\bibitem[{{Lis} \& {Goldsmith}(1990)}]{Goldsmith90}
{Lis}, D.~C., \& {Goldsmith}, P.~F. 1990, \apj, 356, 195

\bibitem[{{Lodders}(2003)}]{Lodders03}
{Lodders}, K. 2003, \apj, 591, 1220

\bibitem[{{Morris} \& {Serabyn}(1996)}]{Morris96}
{Morris}, M., \& {Serabyn}, E. 1996, \araa, 34, 645

\bibitem[{{Muno} {et~al.}(2007){Muno}, {Baganoff}, {Brandt}, {Park}, \&
  {Morris}}]{Muno07}
{Muno}, M.~P., {Baganoff}, F.~K., {Brandt}, W.~N., {Park}, S., \& {Morris},
  M.~R. 2007, \apjl, 656, L69

\bibitem[{{Murakami} {et~al.}(2001){Murakami}, {Koyama}, \&
  {Maeda}}]{Murakami01}
{Murakami}, H., {Koyama}, K., \& {Maeda}, Y. 2001, \apj, 558, 687

\bibitem[{{Murakami} {et~al.}(2000){Murakami}, {Koyama}, {Sakano}, {Tsujimoto},
  \& {Maeda}}]{Murakami00}
{Murakami}, H., {Koyama}, K., {Sakano}, M., {Tsujimoto}, M., \& {Maeda}, Y.
  2000, \apj, 534, 283

\bibitem[{{Nakajima} {et~al.}(2009){Nakajima}, {Tsuru}, {Nobukawa},
  {Matsumoto}, {Koyama}, {Murakami}, {Senda}, \& {Yamauchi}}]{Nakajima09}
{Nakajima}, H., {Tsuru}, T.~G., {Nobukawa}, M., {Matsumoto}, H., {Koyama}, K.,
  {Murakami}, H., {Senda}, A., \& {Yamauchi}, S. 2009, \pasj, 61, 233

\bibitem[{{Oka} {et~al.}(2005){Oka}, {Geballe}, {Goto}, {Usuda}, \&
  {McCall}}]{Oka05}
{Oka}, T., {Geballe}, T.~R., {Goto}, M., {Usuda}, T., \& {McCall}, B.~J. 2005,
  \apj, 632, 882

\bibitem[{{Park} {et~al.}(2004){Park}, {Muno}, {Baganoff}, {Maeda}, {Morris},
  {Howard}, {Bautz}, \& {Garmire}}]{Park04}
{Park}, S., {Muno}, M.~P., {Baganoff}, F.~K., {Maeda}, Y., {Morris}, M.,
  {Howard}, C., {Bautz}, M.~W., \& {Garmire}, G.~P. 2004, \apj, 603, 548

\bibitem[{{Pellegrini} {et~al.}(2000){Pellegrini}, {Cappi}, {Bassani},
  {Malaguti}, {Palumbo}, \& {Persic}}]{Pellegrini00}
{Pellegrini}, S., {Cappi}, M., {Bassani}, L., {Malaguti}, G., {Palumbo},
  G.~G.~C., \& {Persic}, M. 2000, \aap, 353, 447

\bibitem[{{Ponti} {et~al.}(2010)}]{Ponti10}
{Ponti}, G., {et~al.} 2010, submitted to ApJ

\bibitem[{{Porquet} {et~al.}(2008){Porquet}, {Grosso}, {Predehl}, {Hasinger},
  {Yusef-Zadeh}, {Aschenbach}, {Trap}, {Melia}, {Warwick}, {Goldwurm},
  {B{\'e}langer}, {Tanaka}, {Genzel}, {Dodds-Eden}, {Sakano}, \&
  {Ferrando}}]{Porquet08}
{Porquet}, D., {et~al.} 2008, \aap, 488, 549

\bibitem[{{Protheroe} {et~al.}(2008){Protheroe}, {Ott}, {Ekers}, {Jones}, \&
  {Crocker}}]{Protheroe08}
{Protheroe}, R.~J., {Ott}, J., {Ekers}, R.~D., {Jones}, D.~I., \& {Crocker},
  R.~M. 2008, \mnras, 390, 683

\bibitem[{{Quarles}(1976)}]{Quarles76}
{Quarles}, C.~A. 1976, \pra, 13, 1278

\bibitem[{{Reid} {et~al.}(2009){Reid}, {Menten}, {Zheng}, {Brunthaler}, \&
  {Xu}}]{Reid09}
{Reid}, M.~J., {Menten}, K.~M., {Zheng}, X.~W., {Brunthaler}, A., \& {Xu}, Y.
  2009, \apj, 705, 1548

\bibitem[{{Renaud} {et~al.}(2006){Renaud}, {Gros}, {Lebrun}, {Terrier},
  {Goldwurm}, {Reynolds}, \& {Kalemci}}]{Renaud06}
{Renaud}, M., {Gros}, A., {Lebrun}, F., {Terrier}, R., {Goldwurm}, A.,
  {Reynolds}, S., \& {Kalemci}, E. 2006, \aap, 456, 389

\bibitem[{{Revnivtsev} {et~al.}(2004){Revnivtsev}, {Churazov}, {Sazonov},
  {Sunyaev}, {Lutovinov}, {Gilfanov}, {Vikhlinin}, {Shtykovsky}, \&
  {Pavlinsky}}]{Revnivtsev04}
{Revnivtsev}, M.~G., {et~al.} 2004, \aap, 425, L49

\bibitem[{{Ryu} {et~al.}(2009){Ryu}, {Koyama}, {Nobukawa}, {Fukuoka}, \&
  {Tsuru}}]{Ryu09}
{Ryu}, S.~G., {Koyama}, K., {Nobukawa}, M., {Fukuoka}, R., \& {Tsuru}, T.~G.
  2009, \pasj, 61, 751

\bibitem[{{Sakano} {et~al.}(2002){Sakano}, {Koyama}, {Murakami}, {Maeda}, \&
  {Yamauchi}}]{Sakano02}
{Sakano}, M., {Koyama}, K., {Murakami}, H., {Maeda}, Y., \& {Yamauchi}, S.
  2002, \apjs, 138, 19

\bibitem[{{Strong} {et~al.}(2000){Strong}, {Moskalenko}, \&
  {Reimer}}]{Strong00}
{Strong}, A.~W., {Moskalenko}, I.~V., \& {Reimer}, O. 2000, \apj, 537, 763

\bibitem[{{Sunyaev} \& {Churazov}(1998)}]{Sunyaev98}
{Sunyaev}, R., \& {Churazov}, E. 1998, \mnras, 297, 1279

\bibitem[{{Sunyaev} {et~al.}(1993){Sunyaev}, {Markevitch}, \&
  {Pavlinsky}}]{Sunyaev93}
{Sunyaev}, R.~A., {Markevitch}, M., \& {Pavlinsky}, M. 1993, \apj, 407, 606

\bibitem[{{Takagi} {et~al.}(2002){Takagi}, {Murakami}, \& {Koyama}}]{Takagi02}
{Takagi}, S.-i., {Murakami}, H., \& {Koyama}, K. 2002, \apj, 573, 275

\bibitem[{{Tatischeff} {et~al.}(1998){Tatischeff}, {Ramaty}, \&
  {Kozlovsky}}]{Tatischeff98}
{Tatischeff}, V., {Ramaty}, R., \& {Kozlovsky}, B. 1998, \apj, 504, 874

\bibitem[{{Valinia} {et~al.}(2000){Valinia}, {Tatischeff}, {Arnaud}, {Ebisawa},
  \& {Ramaty}}]{Valinia00}
{Valinia}, A., {Tatischeff}, V., {Arnaud}, K., {Ebisawa}, K., \& {Ramaty}, R.
  2000, \apj, 543, 733

\bibitem[{{Verner} \& {Yakovlev}(1995)}]{Verner95}
{Verner}, D.~A., \& {Yakovlev}, D.~G. 1995, \aaps, 109, 125

\bibitem[{{Winkler} {et~al.}(2003){Winkler}, {Courvoisier}, {Di Cocco},
  {Gehrels}, {Gim{\'e}nez}, {Grebenev}, {Hermsen}, {Mas-Hesse}, {Lebrun},
  {Lund}, {Palumbo}, {Paul}, {Roques}, {Schnopper}, {Sch{\"o}nfelder},
  {Sunyaev}, {Teegarden}, {Ubertini}, {Vedrenne}, \& {Dean}}]{Winkler03}
{Winkler}, C., {et~al.} 2003, \aap, 411, L1

\bibitem[{{Yusef-Zadeh} {et~al.}(2002){Yusef-Zadeh}, {Law}, \&
  {Wardle}}]{Yusef02}
{Yusef-Zadeh}, F., {Law}, C., \& {Wardle}, M. 2002, \apjl, 568, L121

\bibitem[{{Yusef-Zadeh} {et~al.}(2007){Yusef-Zadeh}, {Muno}, {Wardle}, \&
  {Lis}}]{Yusef07}
{Yusef-Zadeh}, F., {Muno}, M., {Wardle}, M., \& {Lis}, D.~C. 2007, \apj, 656,
  847

\end{thebibliography}

\end{document}